\documentclass[%
 reprint,
 amsmath,amssymb,
 aps,
 pra,
]{revtex4-1}

\usepackage{graphicx}
\usepackage{dcolumn}
\usepackage{bm}
\usepackage{nicefrac}

\usepackage{color}

\begin{document}

\preprint{APS/123-QED}


\title{Two symmetric four-wave mixing signals generated in a medium with anomalous refractive index}

\author{A. S. Alvarez}

\author{A. A. C. de Almeida}

\author{S. S. Vianna}%
 \email{vianna@ufpe.br}
\affiliation{Departamento de F\'{i}sica, Universidade Federal de Pernambuco, 50670-901, Recife, Pernambuco, Brazil}

\date{\today}

\begin{abstract}
We report experimental and theoretical results of two symmetrical signals of degenerate four-wave mixing generated in rubidium vapor. Both nonlinear signals are induced by two almost copropagating laser beams, with  $\vec{k}_{a}$ and $\vec{k}_{b}$ wave-vectors, and detected simultaneously in the $2\vec{k}_{a}-\vec{k}_{b}$ and $2\vec{k}_{b}-\vec{k}_{a}$ directions. In each direction, we observe a single peak when the two beams are tuned on the closed transition $^{85}$Rb$~5S_{1/2} (F=3) \rightarrow 5P_{3/2} (F=4)$. The excitation spectra reveal a small frequency separation between the two peaks, which is explained when propagation effects are taken into account. Furthermore, our theoretical analysis shows that a correct description of the frequency position of each peak is obtained only if coherent effects such as electromagnetically induced absorption are included.
\end{abstract}

\maketitle

\section{Introduction}
In a four-wave mixing (FWM) process a fourth field is generated as the result of the coherent combination of three electromagnetic fields interacting with a nonlinear sample. This process has been used extensively to investigate a variety of optical phenomena in atomic systems. Since very early studies, different atomic level configurations as, for instance, two-level \cite{abrams1978degenerate, oria1989efficient}, three-level $\Lambda$ \cite{pinard1987backward, cardoso2002electromagnetically}, and four-level double-$\Lambda$ schemes \cite{lukin2000resonant} have been explored to enhance the efficiency of this nonlinear process. One common characteristic in much of these FWM process is the possibility to control the refractive index \cite{lukin1998resonant} of the medium, and in some conditions cancel the resonant absorption due to the phenomenon of electromagnetically induced transparency (EIT) \cite{harris1997electromagnetically, fleischhauer2005electromagnetically} or enhance the absorption via electromagnetically induced absorption (EIA) \cite{akulshin1998electromagnetically, taichenachev1999electromagnetically}. Another interesting feature is the generation of narrowband photon pairs in atomic ensembles via spontaneous FWM \cite{kolchin2006generation} as well as the efficient generation of pairs of intense light beams showing a high degree of intensity squeezing \cite{boyer2007ultraslow,mccormick2007strong}. 

The present work is concerned with two symmetric FWM signals that are generated together in a sample of thermal rubidium atoms. The nonlinear signals are induced by two independent laser beams, with $\vec{k}_{a}$ and $\vec{k}_{b}$ wave-vectors, when both beams are tuned on the same $^{85}$Rb Doppler line $~5S_{1/2} (F=3) \rightarrow 5P_{3/2}$. In such a case, where all fields are almost resonant with the same optical transition, the degenerate FWM signals have been analyzed considering a pure \cite{boyd1981four,steel1981multiresonant} or degenerate two-level system \cite{lipsich2000absorption}, with one strong field, and arbitrary polarization of the drive fields \cite{akulshin2000highly,lezama2000polarization}.

Most of these experiments are performed with a counterpropagating beam configuration, exploring the phase-matching obtained when the generated beam is phase-conjugated with the probe beam. In particular, we employ a copropagating beam configuration and detect simultaneously the transmission of the incident beams and the generated FWM signals at directions $2\vec{k}_{a}-\vec{k}_{b}$ and $2\vec{k}_{b}-\vec{k}_{a}$. In each direction, the excitation spectra show a single FWM peak when the two beams are tuned on the closed transition $^{85}$Rb$~5S_{1/2} (F=3) \rightarrow 5P_{3/2} (F=4)$. A similar scheme, where two FWM fields are also detected simultaneously, has explored a non-degenerate system and showed that the generated fields with different frequencies, like Stoke and anti-Stokes, can be correlated or anticorrelated depending on the incident beams \cite{yang2012generation}. In the experiment described here, the degeneracy of the nonlinear process in combination with incident beams of the same intensities leads to two symmetric signals, either in space as in frequency, independent of which beam is scanning. 

It is interesting to note that although the two signals are generated by two independent FWM processes, they give information about the dynamic of an ensemble of atoms that interacted simultaneously with the same drive fields. Actually, for a spatially uniform atomic medium, the coherent superposition of the generated fields at different positions along the nonlinear medium leads to the well-known phase-matching condition. This condition determines not only the propagation direction of the outgoing FWM field in terms of the wave-vectors of the incident waves but also the frequencies at which the signals will be maximal. Recent studies \cite{zhou2018influence} in a non-degenerated three-level system, show that the phase-matching condition is responsible for the high efficient FWM signal when the excitation fields are turned off-resonance from the atomic transition. In this case, with counter-propagating beams, the predominant contributions are attributed to the EIA grating effects \cite{zhang2011enhanced}. 

The experiment described here, with co-propagating beams and involving a degenerate process, also reveals a frequency shift out of the resonance. In special, we observe a frequency separation between the peaks associated with each one of the FWM signals, with a red or blue frequency shift depending on the relation between the observed signal and the beam which frequency is scanning. This work aims to investigate the main physical mechanisms responsible for these frequency shifts and how they are related to the coherence induced in the atomic system.  In Sec II, the experimental setup and the principal results are presented. Sec. III is devoted to describe the theoretical model and discuss the different contributions to the FWM spectra calculated for a Doppler-broadened sample. We conclude by summarizing the most important results in Sec. IV.

\section{Experimental setup and results}

A simplified scheme of the experimental setup is presented in Fig. 1(a) together with the hyperfine structure of the $D_{2}$ line of $^{85}$Rb. Two independent \textit{cw} diode lasers generate the two beams $E_{a}$ and $E_{b}$ that are responsible for driving the four-wave mixing process. The beams $E_{a}$ and $E_{b}$, with wave-vectors $\vec{k}_{a}$ and $\vec{k}_{b}$, respectively, and orthogonal and linear polarization, converge inside a 5 cm long cell containing a natural concentration of rubidium atoms at an angle of approximately 40 mrad. To increase the atomic density, the rubidium cell was heated to $\sim$ 55 $^{o}$C. Both beams are tuned on the same Doppler line of $^{85}$Rb starting in the hyperfine ground state $F = 3$, as shown in the inset of Fig. 1(a). To control and monitor the frequency of each laser we use saturated absorption spectroscopy (not shown).

We simultaneously detect four signals: the transmission of both incident beams $T_{a}$ and $T_{b}$ and the two FWM signals in the $2\vec{k}_{a}-\vec{k}_{b}$ and $2\vec{k}_{b}-\vec{k}_{a}$ directions, as shown in Fig. 1(b). In this type of forward geometry, the clean detection of the signals can be a challenge since scattered light from one beam might interfere with another's detection. To deal with this we take advantage of the linear and cross-polarization of the signals and use polarizing beamsplitters before each detector. The incident beams are typically strong allowing the use of regular photo-diode detectors to acquire $T_{a}$ and $T_{b}$. On the other hand, the FWM signals are very weak so we use avalanche photo-diodes (\textit{Thorlabs APD120A}) to detect them.

\begin{figure}[htbp]
\centering
\includegraphics[width=1\linewidth, angle=0]{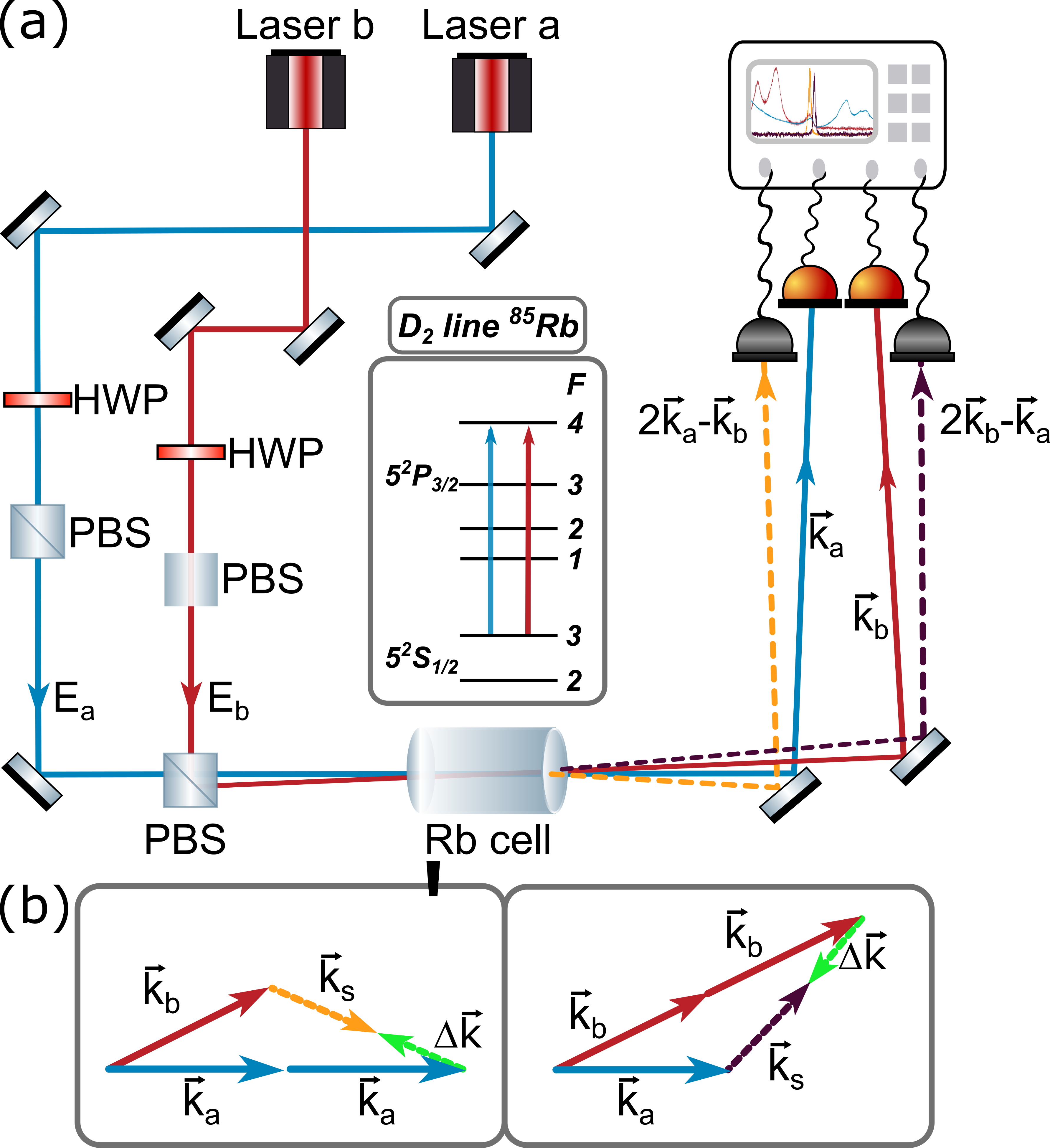}
\caption{(a) Experimental setup with relevant energy levels of $^{85}$Rb. PBS and HWP indicate polarizing beamsplitter and half wave-plate, $\vec{k}_{a}$ and $\vec{k}_{b}$ are the wave-vectors of the $E_{a}$ and $E_{b}$ beams and $2\vec{k}_{a}-\vec{k}_{b}$ and $2\vec{k}_{b}-\vec{k}_{a}$ indicate the directions of two detected FWM signals. (b) Schematic representation of phase-matching for the generation of the two FWM signals.}
\label{fig1}
\end{figure}

The measurements are performed scanning the frequency of one laser throughout the three allowed hyperfine transitions while the other has a fixed frequency. The intensity of the incident lasers at the cell entrance is ten to a hundred times the saturation intensity of the cyclic transition $5S_{1/2} (F=3) \rightarrow 5P_{3/2} (F=4)$. Naturally, absorption is increased due to the temperature of the vapor but not enough to absorb completely the fields $E_{a}$ and $E_{b}$, so we consider to be operating in a high-intensity regime.

A typical experimental result is shown in Fig. 2(a) with the transmission of the two beams $T_{a}$ and $T_{b}$ and the two generated FWM signals, $2\vec{k}_{a}-\vec{k}_{b}$ and $2\vec{k}_{b}-\vec{k}_{a}$, as a function of the detuning ($\delta_{a}/2\pi$) of the scanning field $E_{a}$. The intensities of $E_{a}$ and $E_{b}$ at the cell entrance were selected to be approximately the same ($I\sim$50 mW/cm$^{2}$) and the frequency of $E_{b}$ beam was fixed near the center of the Doppler broadened spectrum. All curves are independently normalized and we chose to measure the frequency detuning with respect to the $5S_{1/2} (F=3) \rightarrow 5P_{3/2} (F=4)$ closed transition.

\begin{figure}[htbp]
\centering
\includegraphics[width=1\linewidth, angle=0]{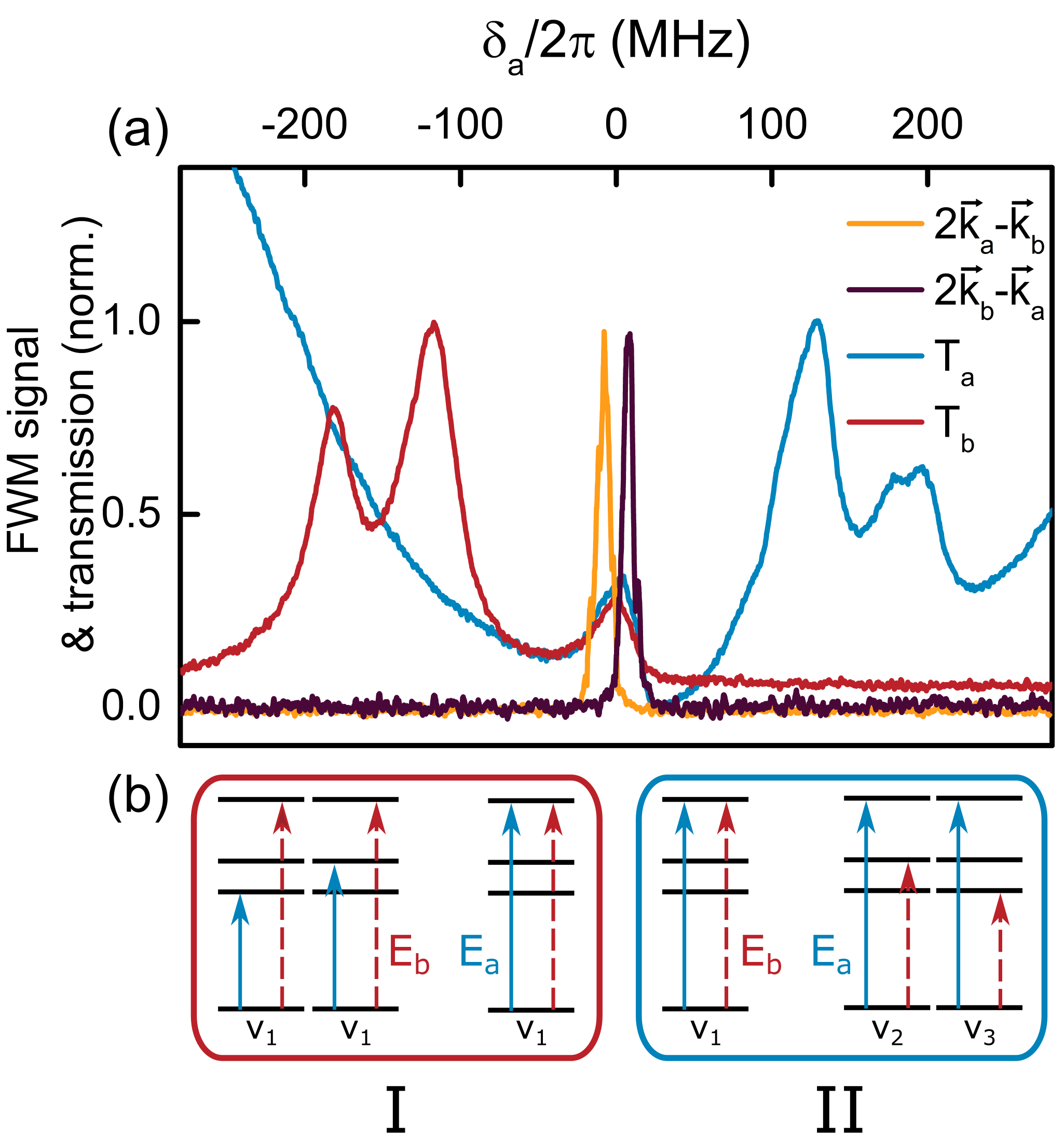}
\caption{(a) Experimental result presenting the transmission of the two incident beams and the two generated FWM signals as a function of $\delta_{a}$. (b) Boxes I and II show the transitions from the ground state $^{85}$Rb $F = 3$, involved on each peak of the transmission signals.}
\label{fig2}
\end{figure}

This result contains a series of interesting features. To begin the analysis with the transmission signals, notice that there are three peaks in each curve. The frequency difference between them reveals that they are related to the hyperfine transitions from the ground state $F = 3$ of $^{85}$Rb. Furthermore, these peaks only appear in the spectra due to the simultaneous interaction of the two incident fields with the atomic medium. These sub-Doppler peaks have been studied when the transmission of a weak beam is measured in the presence of a strong fixed frequency field, and the effect is known as velocity selective optical pumping \cite{moon2008analytic}. Here, we measure the transmission of the two beams, when both are strong.

Notice that the peaks at the transmission curve of the laser with a fixed frequency, $T_{b}$, appears in the appropriate order of energy growth. The same is not true for the transmission of the scanning laser. To explain this we must look at the different atomic velocity groups that interact with the lasers. The field $E_{b}$ induces the closed transition $~5S_{1/2} (F=3) \rightarrow 5P_{3/2} (F=4)$ for a group of atoms with velocity $v_{1}$, as the dashed lines indicate in the box I of Fig. 2(b). As we scan the frequency of $E_{a}$, it executes each one of the allowed transitions, resulting in a smaller absorption of the field $E_{b}$ at these specific frequencies, creating the peaks on the $T_{b}$ curve. 

As for the transmission $T_{a}$, the field $E_{b}$ now selects three groups of atoms with velocities $v_{1}$, $v_{2}$ and $v_{3}$, promoting the transition to the excited states $F=4$, $F=3$ and $F=2$, respectively. Once again, as we scan $E_{a}$, it will have the resonance frequency of the closed transition, as the box II indicates in Fig. 2(b). For each velocity group, the optical pumping due to $E_{b}$ lowers the absorption of $E_{a}$ and generates the three peaks in the opposite order of energy growth \cite{kim2003observation, moon2008analytic, garcia2018velocity}.

The FWM signal we obtain is due to a degenerate process when both fields interact with the velocity group $v_{1}$. In this case, as Fig. 2(b) shows the two incident fields induce the closed transition to the excited state $F = 4$. Notice that the two FWM signals in Fig. 2(a)  present a small frequency separation. One would expect that since both processes are nearly identical, the output signals should not have different positions in the spectrum. As we discuss in the following section, such features are a result of the phase-matching conditions together with coherent effects such as EIA. 

To highlight the presence of the EIA phenomenon we present one of the transmission curves alongside the two FWM signals in Fig. 3. In this case, we use an experimental configuration where the ratio between the intensities of the incident beams is about three, with $E_{a}$ being more powerful. In Fig. 3(a), we scan the frequency of the field $E_{a}$ while $E_{b}$ has a fixed frequency and vice versa for Fig. 3(b). 

\begin{figure}[htbp]
\centering
\includegraphics[width=1\linewidth, angle=0]{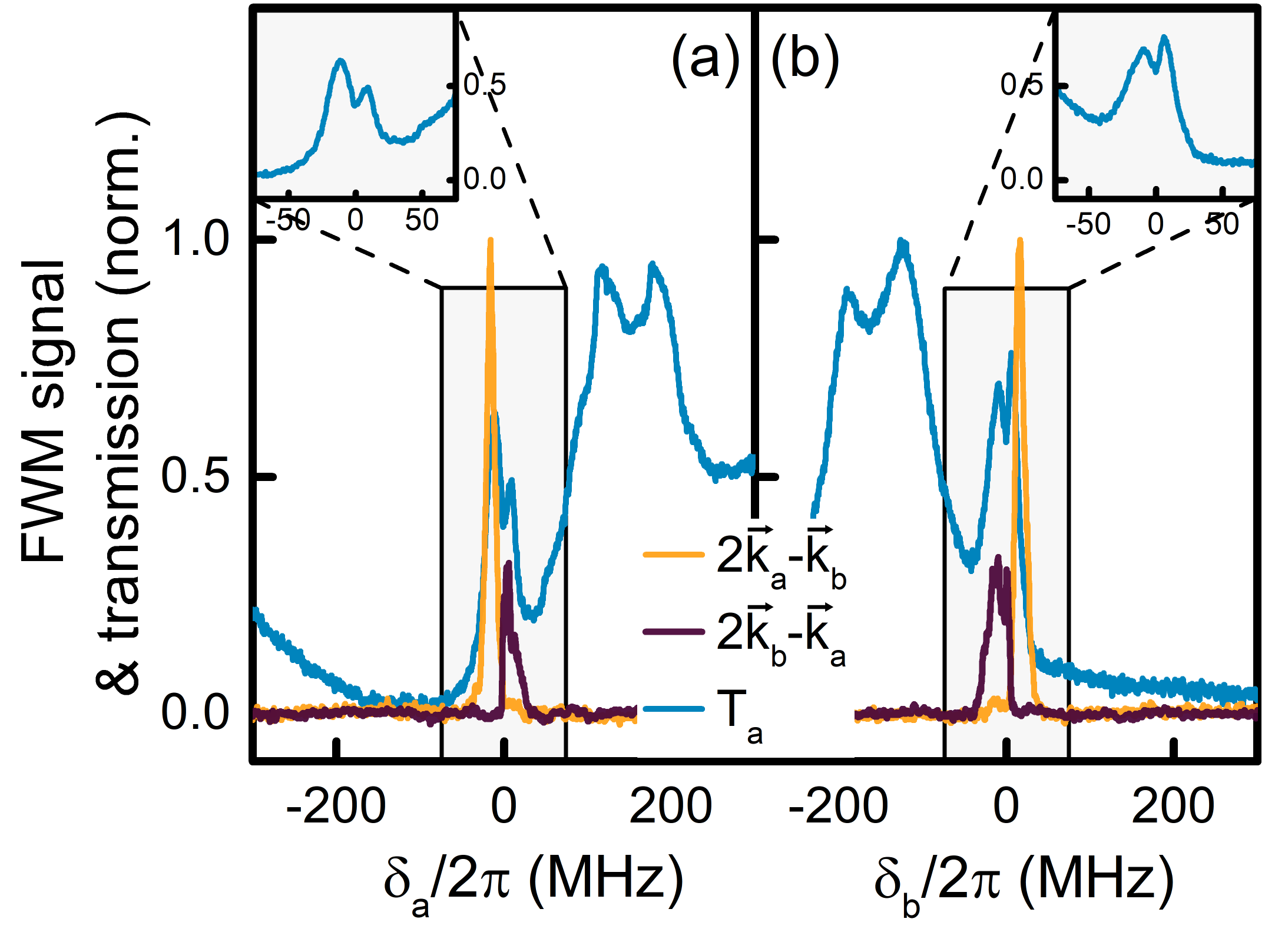}
\caption{Measurements of T$_{a}$ and FWM signals as a function of the frequency detuning of the scanning beam, for a ratio intensity of incident beams of three. (a) Curves for $E_{a}$ scanning and $E_{b}$ fixed in frequency. (b) Curves for $E_{b}$ scanning and $E_{a}$ fixed in frequency. Insets, zoom of the peak corresponding to the cyclic transition on the T$_{a}$ curve in each measurement.}
\label{fig3}
\end{figure}

Since the FWM signals are related to the same incident fields, involving only an interchange of the role of each beam, they present the same intensity relation given by the incident beams. Therefore, the FWM signals are normalized using the highest value between them. As in Fig. 2(a), these generated signals appear again with a small frequency separation. However, we notice an interchange in the frequency positions depending on which beam is scanning. The relative frequency position of the signal $2\vec{k}_{a}-\vec{k}_{b}$ remains the same in relation to the peak corresponding to the cyclic transition in $T_{a}$ curve (see Figs. 2 and 3). The same occurs for the $2\vec{k}_{b}-\vec{k}_{a}$ signal in relation to the peak corresponding to the cyclic transition in $T_{b}$ curve. 

The main feature of Fig. 3 is in the inset: the presence of a narrow absorption dip inside the cyclic transition peak in the $T_{a}$ curve. This absorption dip occurs in the middle of the two nonlinear signals meaning that both lasers are resonant with the closed transition. The narrow dip is the signature of an EIA type process \cite{hossain2011nonlinear} and it can be observed easier when one of the beams is more intense than the other. We use this coherent effect as one of the central pieces in our analysis of the frequency separation of the FWM signals. The EIA dip is also present in the other two peaks of the transmissions curves, although quite smaller and unstable. 

To complete the experimental analysis, we show in Fig. 4 the FWM spectra for four intensities of the incident beams. For these measurements, we scan the frequency of the field $E_{a}$ while $E_{b}$ was fixed in frequency. With the growing intensity of the incident beams, the FWM signals present a power broadening effect together with an increase of the frequency separation. Moreover, for intensities above 200 mW/cm$^{2}$ this frequency separation appears to be saturated. We also observed an asymmetry of the signal that is better defined at higher intensities. 

\begin{figure}[htbp]
\centering
\includegraphics[width=1\linewidth, angle=0]{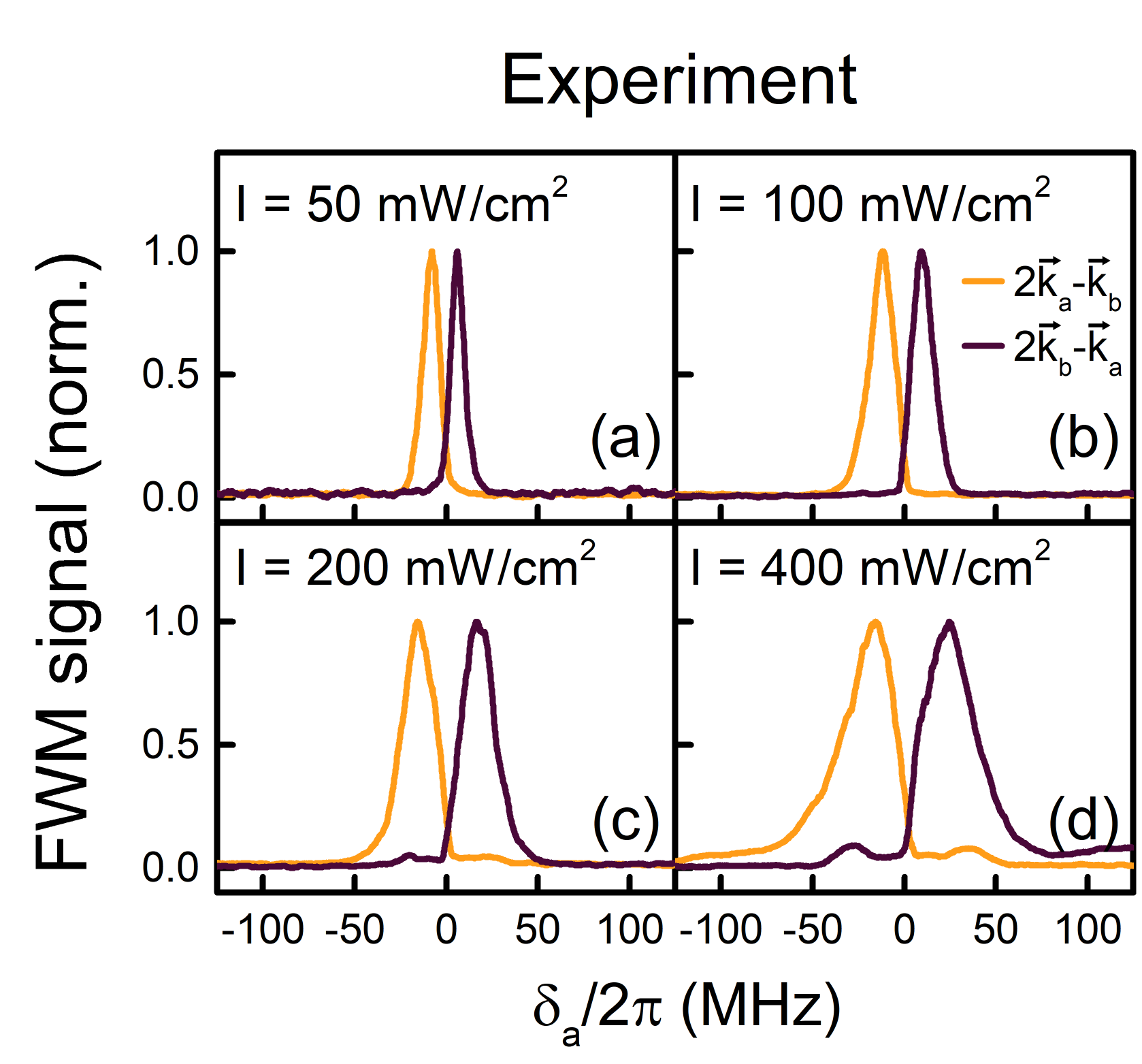}
\caption{Behavior of the two FWM signals at different intensities of the beams $E_{a}$ and $E_{b}$ at the entrance of the Rb cell. All curves are normalized.}
\label{fig4}
\end{figure}

\section{Theory and discussion}

To model our experimental data and explain the main observed features we use a three-level system, as schematized in Fig. 5. We can not use a simpler two-level system since we must take into account the Zeeman structure of the hyperfine levels, accessed by the input lasers due to their orthogonal polarization. Hence,  we model only one process of FWM, while many other similar ones should happen in the different Zeeman sublevels. 

To be consistent with the three-level system, we choose the co-propagation direction of the incident beams as the quantization axis. Therefore, both incident beams will induce $\sigma$ transitions and the generated FWM field will have a circular polarization. However, it is important to note that the degenerescence in the ground state allows a similar and reflected wave mixing to happen simultaneously, generating in phase the other circular component (see Fig. 5(b)) that adds up to a linear polarization. So, in each direction, we detect an FWM signal with linear polarization.

\begin{figure}[htbp]
\centering
\includegraphics[width=1\linewidth]{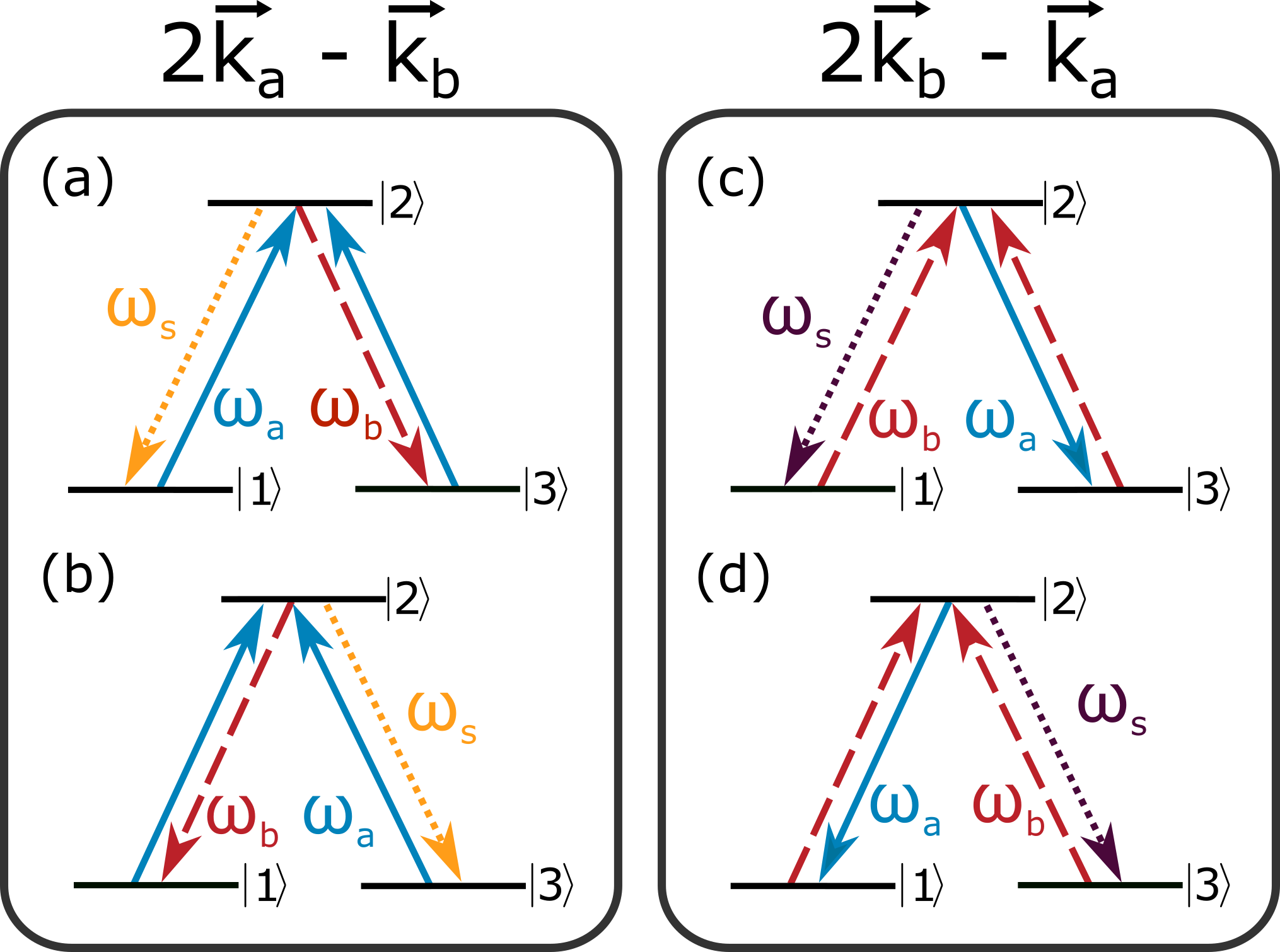}
\caption{(Color online) Three-level theoretical model with (a) and (b) being the processes that generate the two circular components of the signal, with frequency $\omega_{s}$, in the $2\vec{k}_{a}-\vec{k}_{b}$ direction; (c) and (d) are analogous to the $2\vec{k}_{b}-\vec{k}_{a}$ direction.}
\label{fig5}
\end{figure}

Our treatment of the problem begins by considering an electric dipole coupling as the interaction Hamiltonian

\begin{equation}
\hat{H}_{int} = -\hslash\sum^3_{j \neq k}\left(\tilde{\Omega}_{l}e^{i\omega_{l}t-ik_{l}z}+c.c.\right)\left|j\right\rangle\left\langle k\right|,
\end{equation}

\noindent
where $\tilde{\Omega}_{l}=\frac{\mu_{jk}E_{l}}{2\hbar}$ ($l = a$ or $b$) is the Rabi frequency with $\mu_{jk}$ being the transition dipole moment and $E_{l}$ the amplitude of the electric field; $\omega_{l}$ is the optical frequency and $k_{l}$  is the wavenumber associated to the fields indicated in Fig. 5. Using this Hamiltonian it is possible to write Liouville's equation

\begin{equation}
\frac{\partial\rho_{jk}}{\partial t} = -(i\omega_{jk} + \gamma_{jk} + \gamma')\rho_{jk} - \frac{i}{\hslash}\left\langle j \right| [ \hat{H}_{int},\hat{\rho} ] \left| k \right\rangle,\\
\end{equation}

\noindent
where $\gamma_{jk}$ is the decay rate of the density matrix element $\rho_{jk}$, $\gamma'$  is the time of flight decay rate and $\omega_{jk}$ is the frequency of the $\left|j\right\rangle$ $\rightarrow$ $\left|k\right\rangle$ transition. Following the usual steps, we apply the rotating wave approximation and look for a steady-state solution. The Bloch equations for such a system are

\begin{equation}
\begin{split}
\rho_{11}	&=\frac{-i\sigma_{12}\Omega_{a}+i\sigma_{21}\Omega_{a}^{*}+\Gamma_{21}\rho_{22}+\gamma'\rho_{11}^{0}}{\gamma'};\\
\rho_{22}	&=\frac{i\sigma_{12}\Omega_{a}-i\sigma_{21}\Omega_{a}^{*}-i\sigma_{23}\Omega_{b}^{*}+i\sigma_{32}\Omega_{b}}{\Gamma_{21}+\Gamma_{23}+\gamma'};\\
\rho_{33}	&=\frac{i\sigma_{23}\Omega_{b}^{*}-i\sigma_{32}\Omega_{b}+\Gamma_{23}\rho_{22}+\gamma'\rho_{33}^{0}}{\gamma'};\\
\sigma_{12} &=\frac{-i(\rho_{11}-\rho_{22})\Omega_{a}^{*}-i\sigma_{13}\Omega_{b}^{*}}{i\delta_{a}+\gamma_{12}+\gamma'};\\
\sigma_{13}	&=\frac{-i\sigma_{12}\Omega_{b}+i\sigma_{23}\Omega_{a}^{*}}{i\delta_{a}-i\delta_{b}+\gamma_{13}+\gamma'};\\
\sigma_{32}	&=\frac{-i(\rho_{33}-\rho_{22})\Omega_{b}^{*}-i\sigma_{13}\Omega_{a}^{*}}{i\delta_{b}+\gamma_{32}+\gamma'}.\\
\end{split}
\end{equation}

\noindent
The $\sigma_{jk}$ terms are the coherence from Eq. (2) in the rotating frame, whereas $\Gamma_{jk}$ are the decay rates of the populations; the Rabi frequency has been redefined to include the spatial phase $\Omega_{l}=\tilde{\Omega}_{l}e^{-ik_{l}z}$; $\delta_{l}$ is the detuning of each laser with respect to the $\left|j\right\rangle$ $\rightarrow$ $\left|k\right\rangle$ transition; $\rho_{jj}^{0}$ are the populations in the absence of the fields and represent the terms that compensate the loss of atoms from the interaction region with an arrival of new atoms in the ground state at a rate $\gamma'$. The missing coherence equations are the complex conjugate of the ones presented.

We solve the Bloch equations in two stages. In the first, the two fields $E_{a}$ and $E_{b}$ interact with the system to create the coherence $\sigma_{13}$ between the ground states. We carry out this calculation analytically with all the orders of interaction. We may write it as a function of the population terms:

\begin{equation}
\sigma_{13}=\frac{-\Omega_{a}^{*}\Omega_{b}\left[\frac{(\rho_{11}-\rho_{22})}{\left(i\delta_{a}+\gamma_{12}+\gamma'\right)}+\frac{(\rho_{33}-\rho_{22})}{\left(-i\delta_{b}+\gamma_{23}+\gamma'\right)}\right]}{i\delta_{a}-i\delta_{b}+\gamma_{13}+\gamma'+\frac{\left|\Omega_{b}\right|^{2}}{\delta_{a}+\gamma_{12}+\gamma'}+\frac{\left|\Omega_{a}\right|^{2}}{-i\delta_{b}+\gamma_{23}+\gamma'}}.
\end{equation}

\noindent
To find the final expression, i.e., with no dependency on any of the density-matrix elements, we use a computational system with a linear algebra suite. 

In the second stage, we add the interaction with one of the fields, $E_{a}$ or $E_{b}$, in first order to generate the $\sigma'_{21}$ or $\sigma'_{23}$ coherence, responsible for the FWM signals at frequencies $\omega_{s} = 2\omega_{a}-\omega_{b}$ or $2\omega_{b}-\omega_{a}$. In fact, for each direction, both coherence $\sigma'_{21}$ and $\sigma'_{23}$ contribute to the generated signal, each one behind a certain circular component of the signal, as Figs. 5(a) and (b) show.

The joint contribution of the two coherence for the FWM signal in each direction can be equivalently regarded as the scattering of the two circular components of each field $E_{a}$ or $E_{b}$ by the coherence $\sigma_{13}$ created between the two ground states. As the fields co-propagate with a small angle, the scattered fields will travel in different directions. It is useful to notice here that we follow the typical procedure used to describe coherent effects as EIT \cite{fleischhauer2005electromagnetically} and EIA \cite{ taichenachev1999electromagnetically}, by considering all the important contributions to $\sigma_{13}$ that can modify the answer of the medium.

As we can see, both signals generated in the experiment might be modeled with the same set of equations. Therefore, we choose to obtain the equations that describe the $2\vec{k}_{a}-\vec{k}_{b}$ process. We also write the equation for only the circular component of Fig. 5(a), adding both of them at the end of the calculation. In this case, the coherence associate with frequency $\omega_{s}=2\omega_{a}-\omega_{b}$ and wavevector $\vec{k}_{s}=2\vec{k}_{a}-\vec{k}_{b}$ is described by:

\begin{equation}
\sigma'_{21}=\frac{i(\rho_{11}-\rho_{22})\Omega_{s}}{-2i\delta_{a}+i\delta_{b}+\gamma_{12}+\gamma'}+\frac{i\sigma_{31}\Omega_{a}}{-i\delta_{a}+\gamma_{32}+\gamma'}.
\end{equation}

Notice that we include in the first term of Eq. (5) the interaction of the generated field with itself in its lowest order while the second term describes the nonlinear process and is connected directly to the $\sigma_{13}$ coherence. Moreover, the Rabi frequency of the generated signal contains the spatial information, i.e., $\Omega_{s}=\tilde{\Omega}_{s}e^{-ik_{s}z}$.

The FWM signals are, in principle, proportional to the modulus square of the sum $\sigma'_{21} +\sigma'_{23} $. However, this alone would not lead to the asymmetry  and frequency separation observed in the experimental spectra. This response indicates that we must include the propagation of the generated light in the atomic medium. To do so, we use the wave equation obtained from Maxwell's equations 

\begin{equation}
\frac{\partial^{2}E_{s}}{\partial z^{2}}-\frac{1}{n^{2}c^{2}}\frac{\partial^{2}E_{s}}{\partial t^{2}}=\frac{1}{\epsilon_{0}c^{2}}\frac{\partial^{2}P}{\partial t^{2}},
\end{equation}

\noindent
where we assume that the input fields $E_{a}$ and $E_{b}$ are strong enough to allow us to neglect their absorption. $E_{s}$ is the generated electric field and $P=\mu_{12}NTr(\rho)$ \cite{amnon1989quantum} is the macroscopic polarization, with $N$ being the atomic density and $\mu_{12}$ being the dipole moment of the transition. Moreover, the refractive index $n$ for the generated signal is approximated as the index for the field with a fixed frequency. This approximation is to keep consistency with the calculations that follow, although in this particular term $n$ might be considered as equal to unity without great modifications to results.

The polarization has two components $P = P_{linear} + P_{NL}$, each respectively linked to the first and second term on the right-hand side of Eq. (5). We solve the wave equation with a change of variables substituting the electric field $E_{s}$ for the Rabi frequency $\tilde{\Omega}_{s}$, resulting in

\begin{equation}
\tilde{\Omega}_{s}=\frac{\kappa\tilde{\sigma}_{31}\tilde{\Omega}_{a}\left[e^{-\alpha z}-e^{i\Delta kz}\right]}{\left(-i\delta_{a}+\gamma_{32}+\gamma'\right)\left(\alpha+i\Delta k\right)},
\end{equation}

\noindent
where the constant $\kappa = \frac{\omega_{s}N\mu_{12}^{2}}{2\hbar\epsilon_{0}c}$, and the coherence $\tilde{\sigma}_{31}$ is the complex conjugate of Eq. (4) without the spatial dependency in the Rabi frequencies. The absorption coefficient $\alpha$ is taken from Eq. (5)

\begin{equation}
\alpha = \frac{\kappa(\rho_{11}-\rho_{22})}{-2i\delta_{a}+i\delta_{b}+\gamma_{12}+\gamma'}.
\end{equation}

From Fig. 1(b) we may write the phase-matching conditions, $\Delta k = |\Delta \vec{k}|$, for both FWM processes, generated in the $2\vec{k}_{a}-\vec{k}_{b}$ and $2\vec{k}_{b}-\vec{k}_{a}$ directions, respectively, with the field $E_{b}$ with a fixed frequency while we scan the field $E_{a}$

\begin{equation}
\begin{split}
\Delta k_{2\vec{k}_{a}-\vec{k}_{b}}&=\frac{2\omega_{a}}{c}\left\{ n_{a}-n_{b}\cos\left(\theta\right)\right\};\\
\Delta k_{2\vec{k}_{b}-\vec{k}_{a}}&=\frac{1}{c}\left\{ 2\omega_{b}n_{b}\left[1-\cos\left(\theta\right)\right]-\omega_{a}\cos\left(\theta\right)\left[n_{a}-n_{b}\right]\right\}.\\
\end{split}
\end{equation}

\noindent
We write only the phase-matching in the direction of propagation since the angle $\theta$ between input fields is small. As mentioned earlier, we consider the index of refraction for the generated signal as the index of the field with a fixed frequency. 

To model the refractive index we use the real part of the electric susceptibilities of the transition in which laser is. Since these susceptibilities are essentially the coherence that we may extract from the Bloch equations [Eq. (3)], we write the refractive indexes for both beams in the $2\vec{k}_{a}-\vec{k}_{b}$ process as

\begin{equation}
\begin{split}
n_{a}	&=1+\frac{N\mu_{21}^{2}}{2\hbar\epsilon_{0}}\frac{\mathrm{Re}\left(\sigma_{21}\right)}{\Omega_{a}};\\
n_{b}	&=1+\frac{N\mu_{23}^{2}}{2\hbar\epsilon_{0}}\frac{\mathrm{Re}\left(\sigma_{23}\right)}{\Omega_{b}}.
\end{split}
\end{equation}

We obtain these coherences from the previous procedure of solving the nine equations (Eq. (3)) to all orders. The curves for the refractive index as a function of the detuning of the $E_{a}$ field are presented in Fig. 6(a) for a stationary atom. It is important to notice that while $n_{b}$ behaves as one would expect, i.e., it grows with frequency but has an anomalous dispersion around the resonance, $n_{a}$ does not follow the usual behavior. On the other hand, it is interesting that $n_{b}$ is, in fact, a function of the detuning of the $a$ laser, an indicator of the interaction between the fields due to the atomic medium.

It is in the behavior of $n_{a}$ that lies the key factor for the spectral position of the FWM signals. In the place of the anomalous dispersion, there is a window with two of such dispersion and an inflection point on the resonance. This is a typical feature of the EIA process \cite{ling1998electromagnetically}, observed in the transmission of the input beams. We solve the Bloch equations in all orders to reproduce the influence of such a coherent phenomenon on the model.

\begin{figure}[htbp]
\centering
\includegraphics[width=1\linewidth]{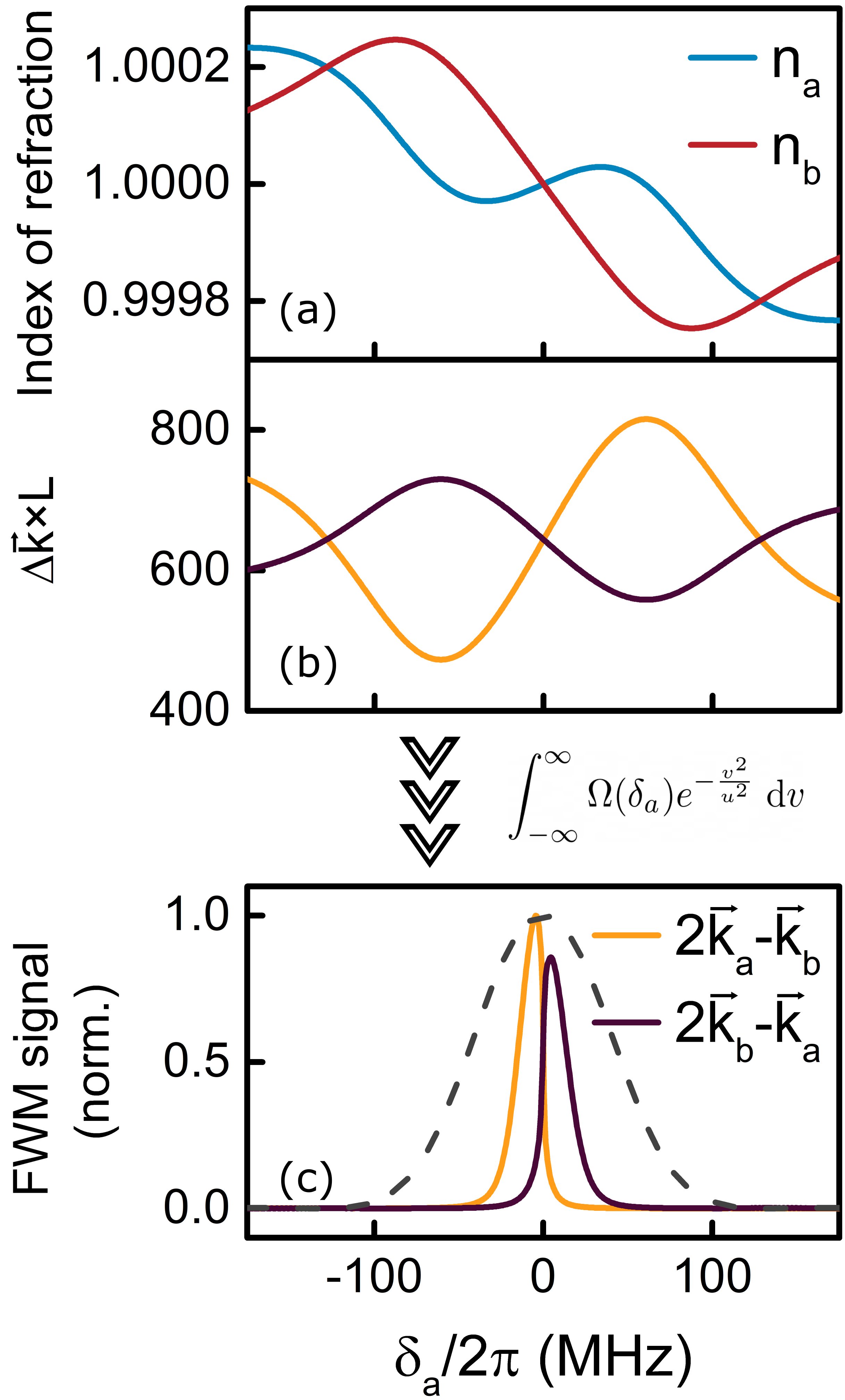}
\caption{(Color online) (a) refractive index for both input fields; (b) Phase-matching conditions for both FWM signals; (c) Theoretical FWM signals with velocity integration and propagation with phase-matching from Eq. 9 (solid line) and with $\Delta k = 0$ (dashed line). All curves are in function of the detuning of the $a$ field. }
\label{fig6}
\end{figure}

With the refractive index modeled, we present the phase-matching condition for both FWM signals in Fig. 6(b), again for an atom with no velocity. Notice that the $\Delta k$ for the $2\vec{k}_{a}-\vec{k}_{b}$ signal is much closer to zero below the resonance and therefore the FWM signal itself should appear on the same region of the spectrum, as presented in the solid curves of Fig. 6(c) [orange/light curve]. The same argument applies to the other signal, $2\vec{k}_{b}-\vec{k}_{a}$ [wine/black curve]. If one considers $\Delta k=0$, i.e. no phase mismatch, the result is shown in the dashed curve of Fig. 6(c). In this case, both FWM spectra are overlapping, broad, and identical.

The modeling of the refractive index that goes in Eq. (10) is critical to understand the features of the experimental signal. If this refractive index behaves as it usually does in a medium with a resonance, i.e., it grows with the frequency with a window an anomalous dispersion around the resonance, then the FWM signal $2\vec{k}_{a}-\vec{k}_{b}$ would only exist above resonance, in complete disagreement with the experiment.

If one chooses to neglect any dispersion effects in the phenomenon, the phase mismatch due to the angle could be obtained, however for a large detuning (of hundreds of MHz).  Therefore, angle alone does not provide the proper $\Delta k$ to be compensated with only the laser detuning. It could be the case if the ground states were not degenerate and therefore, lasers $E_{a}$ and $E_{b}$ had different wavenumbers \cite{zhou2018influence}.

Naturally, one must take into account the Doppler broadening due to the high temperature of the vapor. Consequently, to obtain the FWM spectra in Fig. 6(c) we integrate Eq. (6) with the Maxwell-Boltzmann distribution. In these curves, the atomic density is of the order of $10^{12}$ cm$^{-3}$ and the intensity of both input fields is ten times the saturation intensity of the transition.

The chosen intensity and atomic density are in agreement with the experimental range of the parameters. However, it is important to state that this model does not fit the experimental data perfectly, especially if these two parameters are modified. On the other hand, one can change the intensity, for example, in the experiment and obtain the FWM process for a large set of values. The model itself points in the direction of an explanation to the features of our FWM experimental result. 

The hypothesis supported by our model is, therefore, that the EIA process creates a window in the behavior of the refractive index of the scanning laser. Furthermore, the nonlinear interaction between fields in the medium leads to a variable refractive index for the laser with a fixed frequency. These two effects combine to form the appropriate phase-matching condition to generate one signal below resonance and the other above.

\section{Conclusions}
We have investigated the excitation spectra of two symmetrical FWM signals generated in rubidium vapor, using a copropagating laser beams configuration. The nonlinear signals were induced by two independent lasers when both were tuned on the closed transition $^{85}$Rb$~5S_{1/2} (F=3) \rightarrow 5P_{3/2} (F=4)$, resulting in a single peak in each spectrum. Although this degenerate FWM process is well known, we have detected the two signals simultaneously and explored the symmetry between them. Noteworthy, the results have revealed some anomalies in the index of refraction of the atomic medium induced by the interaction with both fields.

An interesting point is that, even though the two signals are generated by two independent FWM processes, they provide information about the dynamic of an ensemble of atoms that interacted simultaneously with the same excitation fields. In particular, the degeneracy of the nonlinear process in combination with the configuration of the fields leads to a symmetry in the signals, both spatial and in frequency, regardless of which beam is used to probe the excitation spectrum.

The anomalies of the index refraction are unveiled by two experimental features: (i) an  absorption dip in the transmission of the beam that scanning in frequency, like in an EIA process, and (ii) a frequency shift of both FWM signals, in opposite directions, determined by the phase-matching condition.  Our theoretical analysis,  applied to a three-level system, shows how the index of refraction seen by each beam can change during the interaction process. Most importantly, the correct description of the frequency position of each peak is supported by a combination of an EIA process for the scanning laser and a variable refractive index for both lasers.
\vspace{5mm}

This work was supported by CAPES (PROEX 534/2018, No. 23038.003382/2018-39). A. A. C. de Almeida acknowledges financial support by CNPq (141103/2019-1).

\bibliographystyle{ieeetr}

\begin{thebibliography}{10}

\bibitem{abrams1978degenerate}
R.~L. Abrams and R.~C. Lind,  Opt. Lett. \textbf{2}, 94 (1978).

\bibitem{oria1989efficient}
M.~Oria, D.~Bloch, M.~Fichet, and M.~Ducloy, Opt. Lett. \textbf{14}, 1082 (1989).

\bibitem{pinard1987backward}
M.~Pinard, P.~Verkerk, and G.~Grynberg, Phys. Rev. A \textbf{35}, 4679, (1987).

\bibitem{cardoso2002electromagnetically}
G.~C. Cardoso and J.~W.~R. Tabosa  Phys. Rev. A \textbf{65}, 033803 (2002).

\bibitem{lukin2000resonant}
M.~D. Lukin, P.~R. Hemmer, and M.~O. Scully,  Advances in Atomic,
  Molecular, and Optical Physics \textbf{42}, 347 (2000).

\bibitem{lukin1998resonant}
M.~D. Lukin, P.~R. Hemmer, M.~L{\"o}ffler, and M.~O. Scully, Phys. Rev.
  Lett. \textbf{81} 2675 (1998).

\bibitem{harris1997electromagnetically}
S.~E. Harris, Phys. Today \textbf{50}, 36 (1997).

\bibitem{fleischhauer2005electromagnetically}
M.~Fleischhauer, A.~Imamoglu, and J.~P. Marangos, Rev. Mod. Phys. \textbf{77}, 633 (2005).

\bibitem{akulshin1998electromagnetically}
A.~M. Akulshin, S.~Barreiro, and A.~Lezama, Phys. Rev. A \textbf{57}, 2996 (1998).

\bibitem{taichenachev1999electromagnetically}
A.~V. Taichenachev, A.~M. Tumaikin, and V.~I. Yudin, Phys. Rev. A \textbf{61}, 011802(R) (1999).

\bibitem{kolchin2006generation}
P.~Kolchin, S.~Du, C.~Belthangady, G.~Y. Yin, and S.~E. Harris, Phys. Rev.
  Lett. \textbf{97}, 113602 (2006).

\bibitem{boyer2007ultraslow}
V.~Boyer, C.~F. McCormick, E.~Arimondo, and P.~D. Lett, Phys. Rev. Lett.
 \textbf{99}, 143601 (2007).

\bibitem{mccormick2007strong}
C.~F. McCormick, V.~Boyer, E.~Arimondo, and P.~D. Lett, Opt. Lett. \textbf{32}, 178 (2007).

\bibitem{boyd1981four}
R.~W. Boyd, M.~G. Raymer, P.~Narum, and D.~J. Harter, Phys. Rev. A \textbf{24},
  411 (1981).

\bibitem{steel1981multiresonant}
D.~G. Steel and R.~C. Lind, Opt. Lett. \textbf{6}, 587 (1981).

\bibitem{lipsich2000absorption}
A.~Lipsich, S.~Barreiro, A.~M. Akulshin, and A.~Lezama, Phys. Rev. A \textbf{61},
  053803 (2000).

\bibitem{akulshin2000highly}
A.~M. Akulshin, S.~V. Barreiro, and A.~Lezama,  Quantum. Electron. \textbf{30}, 189 (2000).

\bibitem{lezama2000polarization}
A.~Lezama, G.~C. Cardoso, and J.~W.~R. Tabosa,  Phys. Rev. A \textbf{63}, 013805 (2000).

\bibitem{yang2012generation}
X.~Yang, J.~Sheng, U.~Khadka, and M.~Xiao,  Phys. Rev. A \textbf{85}, 013824 (2012).

\bibitem{zhou2018influence}
H.~T. Zhou, S.~N. Che, Y.~H. Han, and D.~Wang,  Indian J. Phys. \textbf{92}, 557 (2018).

\bibitem{zhang2011enhanced}
J.~X. Zhang, H.~T. Zhou, D.~W. Wang, and S.~Y. Zhu, Phys. Rev. A \textbf{83}, 053841 (2011).

\bibitem{moon2008analytic}
G.~Moon and H.~R. Noh, Phys. Rev. A \textbf{78}, 032506 (2008).

\bibitem{kim2003observation}
S.~K. Kim, H.~S. Moon, K.~Kim, and J.~B. Kim,  Phys. Rev. A \textbf{68}, 063813 (2003).

\bibitem{garcia2018velocity}
A.~C. Garc{\'\i}a-Wong, A.~A.~C. de~Almeida, N.~R. de~Melo, and S.~S. Vianna, Opt. Commun. \textbf{427}, 462 (2018).

\bibitem{hossain2011nonlinear}
M.~M. Hossain, S.~Mitra, P.~Poddar, C.~Chaudhuri, B.~Ray, and P.~N. Ghosh, 
  J. Phys. B: At. Mol. Opt. Phys. \textbf{44}, 115501 (2011).

\bibitem{amnon1989quantum}
A.~Yariv, {\em Quantum electronics}.
\newblock Wiley, 1989.

\bibitem{ling1998electromagnetically}
H.~Y. Ling, Y.~Q. Li, and M.~Xiao, Phys. Rev. A \textbf{57}, 1338 (1998).

\end{thebibliography}

\end{document}